\documentclass[useAMS,usenatbib,usegraphicx]{mn2e}

\title[Distances of SNRs Kes 69 and G21.5-0.9]{The Distances of SNRs Kes 69 and G21.5-0.9 from  HI and $^{13}$CO Spectra}

\author[W.W. Tian and D.A. Leahy]{W. W. Tian \thanks{E-mail: tww@iras.ucalgary.ca}, D.A. Leahy \\
Department of Physics $\&$ Astronomy, University of Calgary, Calgary, Alberta T2N 1N4, Canada}

\begin{document}

\date{Accepted Aug. 29 2008. Received 2008; in original form 2008 Aug. 1}

\pagerange{\pageref{firstpage}--\pageref{lastpage}} \pubyear{}

\maketitle

\label{firstpage}
 
\begin{abstract}
We obtain new HI and $^{13}$CO images around Supernova Remnants (SNR) Kes 69 and G21.5-0.9. By comparing HI spectra with $^{13}$CO emission spectra, we significantly revise the kinematic distance for Kes 69 to $\sim$ 5.5 kpc, which was 11.2 kpc, and refine the kinematic distance for G21.5-0.9 to $\sim$ 4.8 kpc. For Kes 69, the highest velocity of absorption is $\sim$ 86 km s$^{-1}$ and a prominent HI emission feature at $\sim$ 112 km s$^{-1}$ has no respective absorption. These new results suggest that Kes 69 is associated with a newly detected extended 1720 MHz OH maser at velocity of $\sim$ 85 km s$^{-1}$ that originates from within the bright southern radio shell of Kes 69.  
For G21.5-0.9, the highest velocity of absorption is $\sim$ 67 km s$^{-1}$. The HI absorption spectra of the nearby bright source PMN J1832-1035 and of Kes 69 show a common absorption feature at velocity of $\sim$ 69 km s$^{-1}$, which is not seen for G21.5-0.9. The resulting velocity of $\sim$ 68 km s$^{-1}$ gives the best distance estimate of $\sim$ 4.8 kpc for G21.5-0.9 and associated young pulsar J1833-1034. 

\end{abstract}
\begin{keywords}
supernova remnants - masers - pulsars:individual:J1833-1034 - cosmic rays - methods:data analysis
\end{keywords}

\section{Background}

Determining distances to Galactic objects (HII regions, pulsars (PSR) and supernova remnants (SNR) etc.) may help understand the kinematics of the Milky Way. Also, distance measurement of SNRs is a key to obtain their basic parameters such as the luminosity, size and age. As an energetic class of objects, SNRs are associated with many highly active astrophysical phenomena, e.g. anomalous X-ray pulsars, soft  $\gamma$-ray repeaters, Pulsar Wind Nebula (PWN), non-thermal X-rays and very-high-energy $\gamma$-rays (see Yang et al. 2008 for a review). The determination of distance of a SNR may test reality of a SNR/PSR/PWN or SNR/TeV $\gamma$-ray source association, and help constrain the mass range of the progenitor star and type of supernova responsible for the remnant. It can provide direct evidence whether a SNR is physically associated with molecular clouds, or if strong interaction between a SNR shock and surrounding clouds is possible, which is widely believed to be a source of TeV $\gamma$-ray and non-thermal X-ray emission in the Milky Way. The SNR/molecular cloud interaction may also potentially lead to new generations of star formation, so constitutes an important part of Galactic ecology. 

Comparing  HI absorption spectra toward  Galactic SNRs with respective HI and $^{13}$CO emission spectra along the line of sight, we previously revised distances to five SNRs and several overlapping HII regions (Tian et al. 2007; Tian \& Leahy 2008; Leahy \& Tian 2008a;). Some values differ greatly from previously published values, e.g. Kes 75 (Leahy \& Tian 2008b), PWN G54.1+0.3 (Leahy, Tian \& Wang 2008), because our methods implement improved background subtraction and spurious emission rejection, and resolve the near/far distance ambiguity in the inner Galaxy, which may have been previously done incorrectly. The distance changes can result in significant changes in interpretation of the SNR (and associated object) properties. As the newest paper of a series, we use the methods to directly re-measure distances of two more SNRs: Kes 69 which has 1720 MHz OH maser emission (Hewitt et al. 2008), and G21.5-0.9 which hosts a young pulsar (Camilo et al. 2006; Bietenholz \& Bartel 2008) and also has TeV $\gamma$-ray emission (i.e. HESS J1833-105, Djannati-Atai et al. 2007). 1720 MHz masers are associated with warm, shocked molecular gas and are seen as signposts of SNR-molecular cloud interactions (Wardle \& Yusef-Zadeh 2002), therefore they likely give reliable distance estimates to SNR/molecular cloud systems.  An OH maser at velocity of 69.3$\pm$0.7 km s$^{-1}$ (hereafter, we use $\sim$ 70 km s$^{-1}$to describe the radial velocity of this maser) was detected toward Kes 69 (Green et al. 1997), leading to a distance of $\sim$ 11.2 kpc for the remnant. However, the masers' site is outside of the northeastern radio and X-ray shells of Kes 69 (Yusef-Zadeh et al. 2003). This spurred us to re-measure its distance by analyzing HI+$^{13}$CO spectra in the direction of Kes 69 in order to test the reality of the claimed association. SNR G21.5-0.9 is an interesting SNR nearby Kes 69, so we also analyze HI and CO spectra in the direction of G21.5-0.9.  

In this paper, we significantly revise the distance to Kes 69, and refine the distance estimation to G21.5-0.9 by analyzing HI and CO spectra. We use the Galactic rotation curve model and recent measurements of the parameters for this (i.e. R$_{0} \sim$ 8 kpc, Eisenhauer et al. 2005; V$_{0}$$\sim$ 220 km s$^{-1}$, Feast \& Whitelock 1997, Reid \& Brunthaler 2004).  The radio data come from 1420 MHz continuum plus HI-line observations of the VLA Galactic Plane Survey (VGPS, Stil et al. 2006) and the $^{13}$CO-line (J = 1-0) observations of the Galactic ring survey (Jackson et al. 2006) of the Five College Radio Astronomy Observatory 14 m telescope. The short-spacing information for the HI spectral line images is from additional observations with the 100 m Green Bank telescope (GBT) of the National Radio Astronomical Observatory (NRAO).

\section{Results}
\subsection{The Radio Continuum and HI Channel Images}
The VLA 1420 MHz continuum image of Kes 69, G21.5-0.9 and background compact source PMN J1832-1035 (i.e. G21.35-0.63) is shown in Fig. 1 (upper left).  We have searched the VGPS radial velocity range from -113 to 165 km s$^{-1}$ for features in the HI emission which might be related to the morphology of the two SNRs.  Some negative HI features are found to be strongly correlated with the continuum intensity of the SNRs, indicating that they are caused by absorption in HI between the SNR and the earth (see the HI channel images shown in the second and third rows of Fig. 1). A CO channel image is also given in Fig. 1 (upper right).

The continuum image clearly reveals that the site of the 1720 MH OH maser at $\sim$ 70 km s$^{-1}$ is outside of the detected radio emission of Kes 69 (marked by the plus in the first panel of Fig. 1).  The CO channel image shows that a cloud at velocity of $\sim$ 86 km s$^{-1}$ overlaps Kes 69, which is likely responsible for producing the clear HI dip in the HI channel image at $\sim$ 86 km s$^{-1}$ at the location of the brightest continuum emission from Kes 69. This means that the CO cloud should be adjacent to or in front of Kes 69. The HI channel image at $\sim$ 112 km s$^{-1}$, which is close to the tangent point velocity of $\sim$ 120 km s$^{-1}$ in the direction towards Kes 69 (see Fig. 2), shows that an HI cloud overlaps Kes 69, but doesn't produce any absorption from Kes 69, so the HI cloud is behind Kes 69.

The HI channel image at $\sim$ 67 km s$^{-1}$ shows that  HI covers Kes 69, G21.5-0.9 and PMN J1832-1035, and produces absorption for each source. So the HI cloud is in front of all three sources. Other HI at $\sim$ 69 km s$^{-1}$ (the lower left panel of Fig. 1) yields an upper limit distance to G21.5-0.9 because it covers all three sources, and produces absorption from Kes 69 and PMN J1832-1035 but not from G21.5-0.9.           

\subsection{The HI and CO Spectra}
We construct HI emission and absorption spectra of both SNRs and of PMN J1832-1035 in order to verify the above results in more detail. The extraction regions for these spectra are shown by the boxes in the first panel of Fig. 1.  The $^{13}$CO emission spectra from boxes 1, 2, 3 and from the OH maser site are also analyzed. These spectra are shown in Fig. 2.  We note that the CO (molecular cloud) emission features all have associated HI emission features indicating HI associated with H$_{2}$, likely due to the atomic envelope of the molecular cloud. Several HI emission features have no associated CO emission (i.e. H$_{2}$), as many atom clouds have no molecular component.

The highest velocity of absorption to Kes 69 is $\sim$ 86 km s$^{-1}$ at which there is a CO cloud (see first and second rows of Fig. 2). The HI spectra of Kes 69 show a prominent HI emission feature at $\sim$ 112 km s$^{-1}$ with no respective absorption. These give a strict distance constraint to Kes 69, i.e. Kes 69 is in the range of distance of 5.5 kpc to 7.4 kpc.   

The $^{13}$CO spectrum of the OH maser site near Kes 69 shows little $^{13}$CO emission at $\sim$ 70 km s$^{-1}$ (above 0.2 K, see the lower right panel), showing there is no high column density cloud. However this cannot exclude there still possibly exists a small high density cloud at the OH maser site at 70 km s$^{-1}$. 

For G21.5-0.9, the highest velocity of HI absorption is $\sim$ 67 km s$^{-1}$ at which there is a CO cloud (third row of Fig. 2). Another CO cloud at $\sim$ 74 km s$^{-1}$ does not produce any respective HI absorption, so this cloud is behind G21.5-0.9. Further, the HI absorption spectra of both nearby PMN J1832-1035 and Kes 69 show a strong absorption feature at velocity of $\sim$ 70 km s$^{-1}$.  These strongly support the above HI channel image analysis and constrain the distance, i.e.  G21.5-0.9 is between  clouds at velocities of $\sim$ 67 km s$^{-1}$ and $\sim$ 69 km s$^{-1}$, i.e. d $\sim$ 4.8 kpc for velocity of $\sim$ 68 km s$^{-1}$.

\section{Discussion}

\subsection{Distances to Kes 69 and G21.5-0.9}

The OH-maser determined distance should be reasonably consistent with the HI+CO determined distance to the same SNR. We note that distances are derived here from a circular rotation model, and contain errors due to uncertainties in the rotation model and non-circular motions. E.g. using the observed $l-V$ diagram, Weiner \& Sellwood (1999) derived the radial velocity distribution in the inner galaxy. Applying to Kes 69, the distances are reduced by $\sim$ 0.4 kpc compared to the circular rotation model. Observed random motion of up to 7 km s$^{-1}$ (Shaveret al. 1982) yields a distance uncertainty of $\sim$ 0.3 kpc for this case. Using the circular rotation model, our analysis of the HI+CO spectra reveals that Kes 69 has a distance of 5.5 to 7.4 kpc, far smaller than 11.2 kpc determined by the OH maser at 70 km s$^{-1}$. This is strong evidence against the OH maser being associated with Kes 69. This is supported by further evidence: the site of the OH maser is outside the detected radio emission of Kes 69, and no CO emission at $\sim$ 70 km s$^{-1}$ is seen in the CO spectrum of the OH maser. A collisional pumping model shows that shock-excited OH maser emission at 1720 MHz appearing behind a SNR shock front results from the interaction of an SNR with an adjacent, warm, dense shocked molecular cloud (Wardle \& Yusef-Zadeh 2002).    

Furthermore, an extended 1720 MHz OH maser at velocity of $\sim$ 85 km s$^{-1}$ has been detected from within the bright southern radio shell of Kes 69 by GBT and VLA observations (Hewitt et al. 2008). The new OH maser fits nicely with the highest velocity of HI absorption in the direction of Kes 69, therefore we conclude that Kes 69 is located at a distance of $\sim$ 5.5 kpc at the near side kinematic distance for the OH maser at $\sim$ 85 km s$^{-1}$ and consistent with the highest observed HI absorption velocity of $\sim$ 86 km s$^{-1}$. A bright $^{13}$CO cloud at $\sim$ 86 km s$^{-1}$ overlapping Kes 69 has been revealed here (Fig. 1 upper right).
A research group from Nan Jing University newly finds $^{12}$CO, $^{13}$ CO and
HCO$^{+}$ emissions near 85 km/s$^{-1}$ from the bright southern shell (Zhou et al. 2008), consistent with the previous formaldehyde molecular (H$_{2}$CO) and the 1665/7 MHz OH absorption line observations at velocities of 82 -- 87 km s$^{-1}$ (Wilson 1972, Turner 1970). All these support our conclusion on the distance of 5.5 kpc to Kes 69.  
So the 1720 MHz maser at 70 km s$^{-1}$ is likely not associated with Kes 69.

G21.5-0.9 is located between a cloud at velocity of $\sim$ 67 km s$^{-1}$
 and that of $\sim$ 69 km s$^{-1}$ (i.e. d $\sim$ 4.8 kpc). G21.5-0.9 hosts a young pulsar PWN J1833-1034. 
New measurement refines the most recent velocity measurement to the SNR/PSR system by Camilo et al. 2006 who analyzed HI spectra data from the VLA and the Leiden/Dwingeloo 25 m telescope observations, and lower resolution $^{12}$CO data from the CfA 1.2 m telescope observations, to obtain lower and upper limits on the velocity of 65 km s$^{-1}$ and 76 km s$^{-1}$ (i.e. d $\sim$ 4.7$\pm$ 0.4 kpc).

\subsection{The Evolutionary State of Kes 69}

The 1420 MHz continuum map shows Kes 69 has a roughly elliptical outline 26$^\prime$ by 20$^\prime$
whereas the ROSAT PSPC image (Yusef-Zadeh et al. 2003), also elliptical and oriented in the same way, has
smaller dimensions of  20$^\prime$ by 13$^\prime$. This is mainly due to absence of X-rays from the
northwest radio filament (upper left side in Fig. 1 here) and from the northeast radio wing (lower left
in Fig. 1), but also the X-rays do not extend out to the edge of the main radio filament in the south. 
We use 20$^\prime$ as the mean angular diameter of Kes 69, which is the mean of radio and X-ray values.
At the distance of 5.5 kpc, the diameter is $\sim$ 32 pc. 

We apply a Sedov model (Cox  1972) to 
estimate the parameters of Kes 69, using the ROSAT PSPC temperature of 1.6 keV and X-ray luminosity of
8.4$\times 10^{34}$ erg/s (Yusef-Zadeh et al. 2003). The high temperature and moderate X-ray
luminosity are indicative of explosion in a low density medium. Application of the Sedov model yields
an age of 5000 years, an explosion energy of 0.8$\times 10^{51}$ erg, and a pre-explosion density of 
$\sim$0.1 cm$^{-3}$. Based on the estimated errors in the ROSAT PSPC spectral parameters, the errors in these 
values are $\sim$60\% for age and explosion energy and $\sim$30\% for density. 
These results are consistent with observation of OH maser emission from the main south filament of
Kes 69 if the explosion occurred in a moderately low density cavity ($\sim$0.1 cm$^{-3}$) and has
only recently ($<<$ 5000 years ago) run into dense molecular gas at the main southern filament. 
Supporting this are Spitzer infrared observations of line emissions from shocked molecular gas in
Kes 69 (Reach et al. 2006).
For the ROSAT temperature of 1.6 keV, the SNR shock velocity is 1250 km s$^{-1}$ in 0.1 cm$^{-3}$ gas, but is 
slowed down to $\sim$ 4 km s$^{-1}$ in cold molecular gas with density of $\sim$ $10^{4}$ cm$^{-3}$. The shock would still
be supersonic, since the sound speed is 0.2 to 0.6 km s$^{-1}$ in 10 to 100 K molecular hydrogen, and
would result in mild heating of the gas to $\sim$ 4000 K. 
This would subsequently result in good conditions for production 
of the observed OH maser emission (Wardle 1999).

In summary, we significantly revise the distance to Kes 69 and obtain a better distance to 
G21.5-0.9, by an analysis using HI and $^{13}$CO spectra. The evolutionary state of Kes 69 is 
evaluated, and it is consistent with a moderate aged SNR, just recently encountering a dense molecular
cloud. 

\begin{figure*}
\vspace{150mm}
\begin{picture}(100,100)
\put(-180,360){\includegraphics{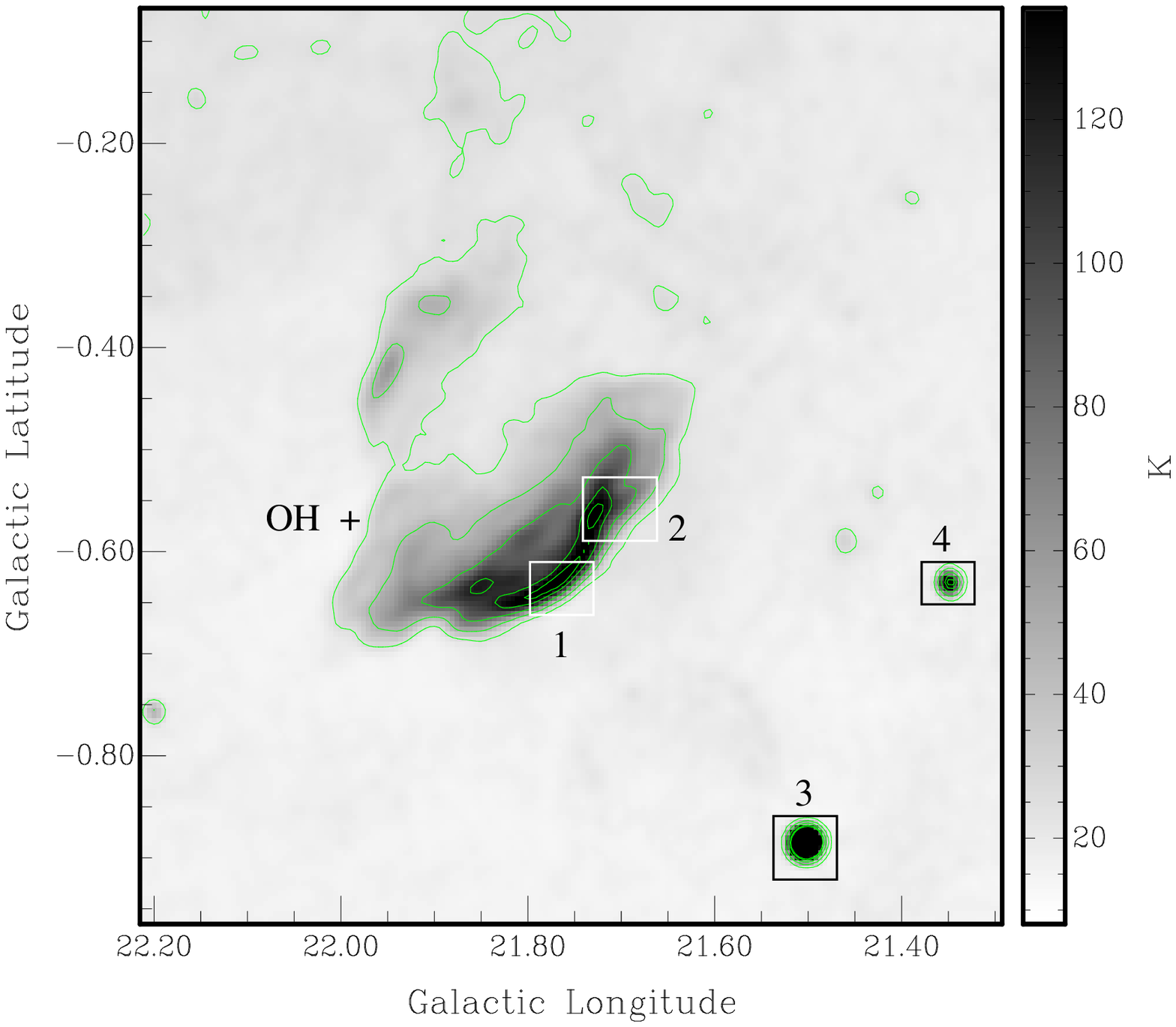}}
\put(-10,590){\includegraphics{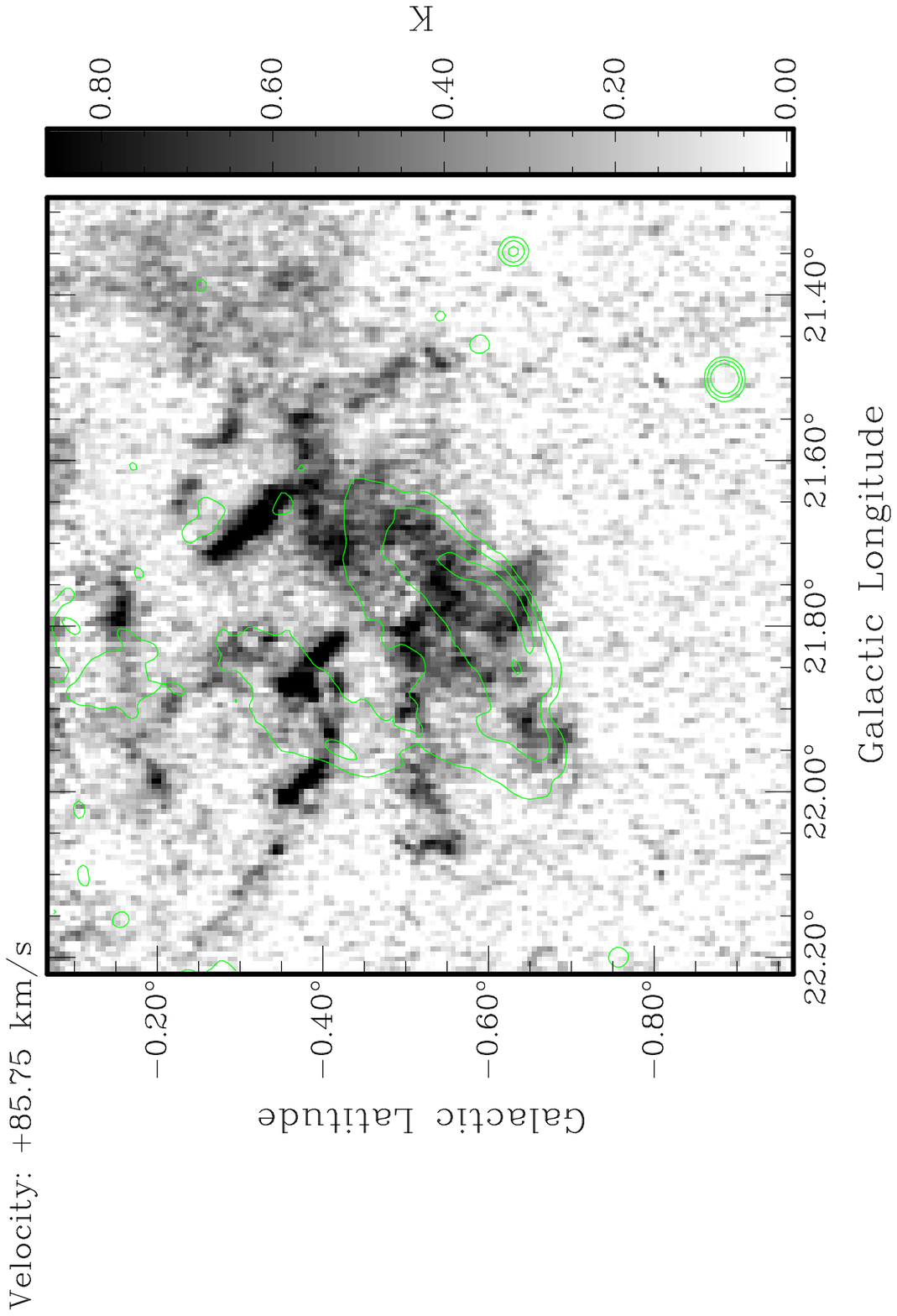}}
\put(-235,385){\includegraphics{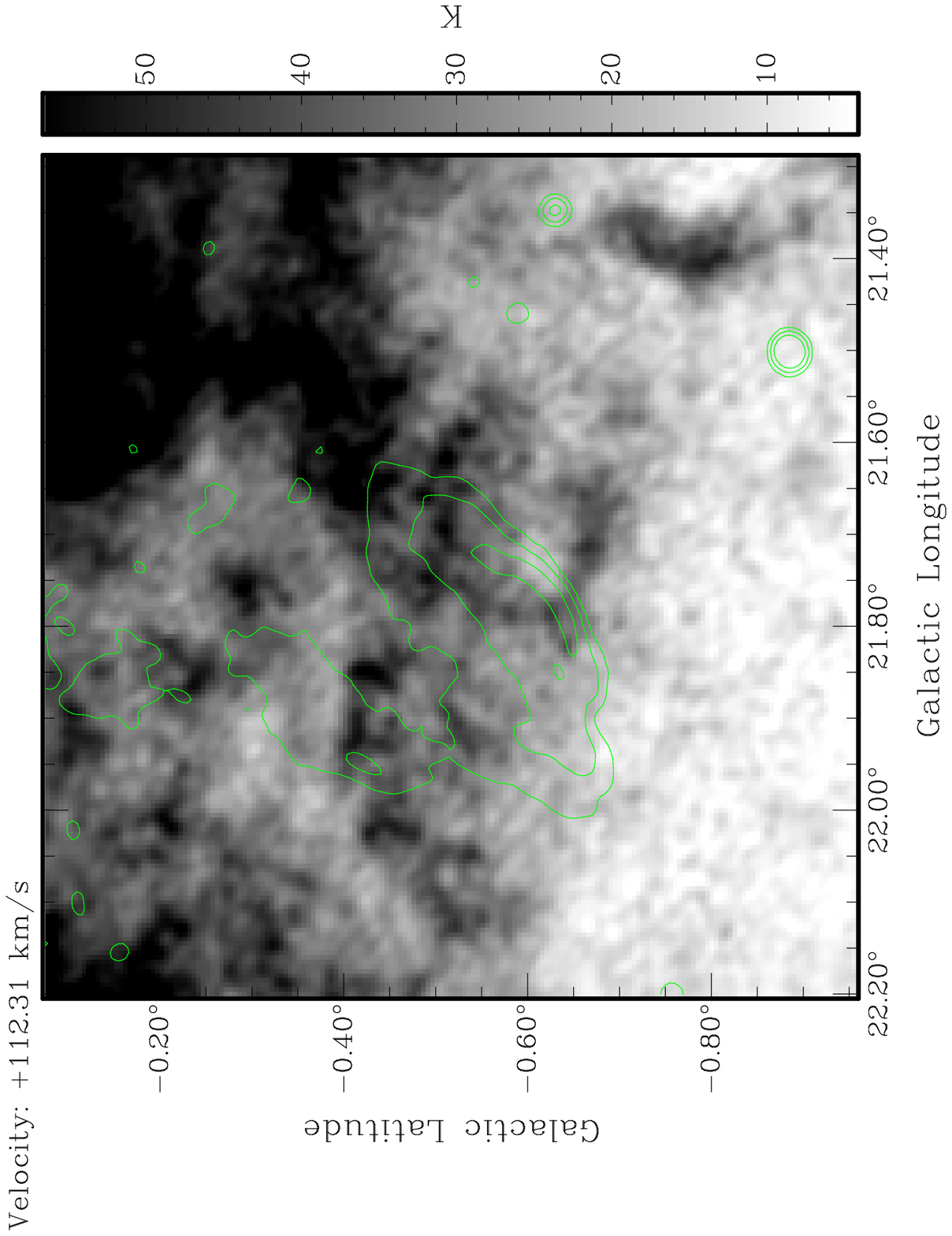}}
\put(30,385){\includegraphics{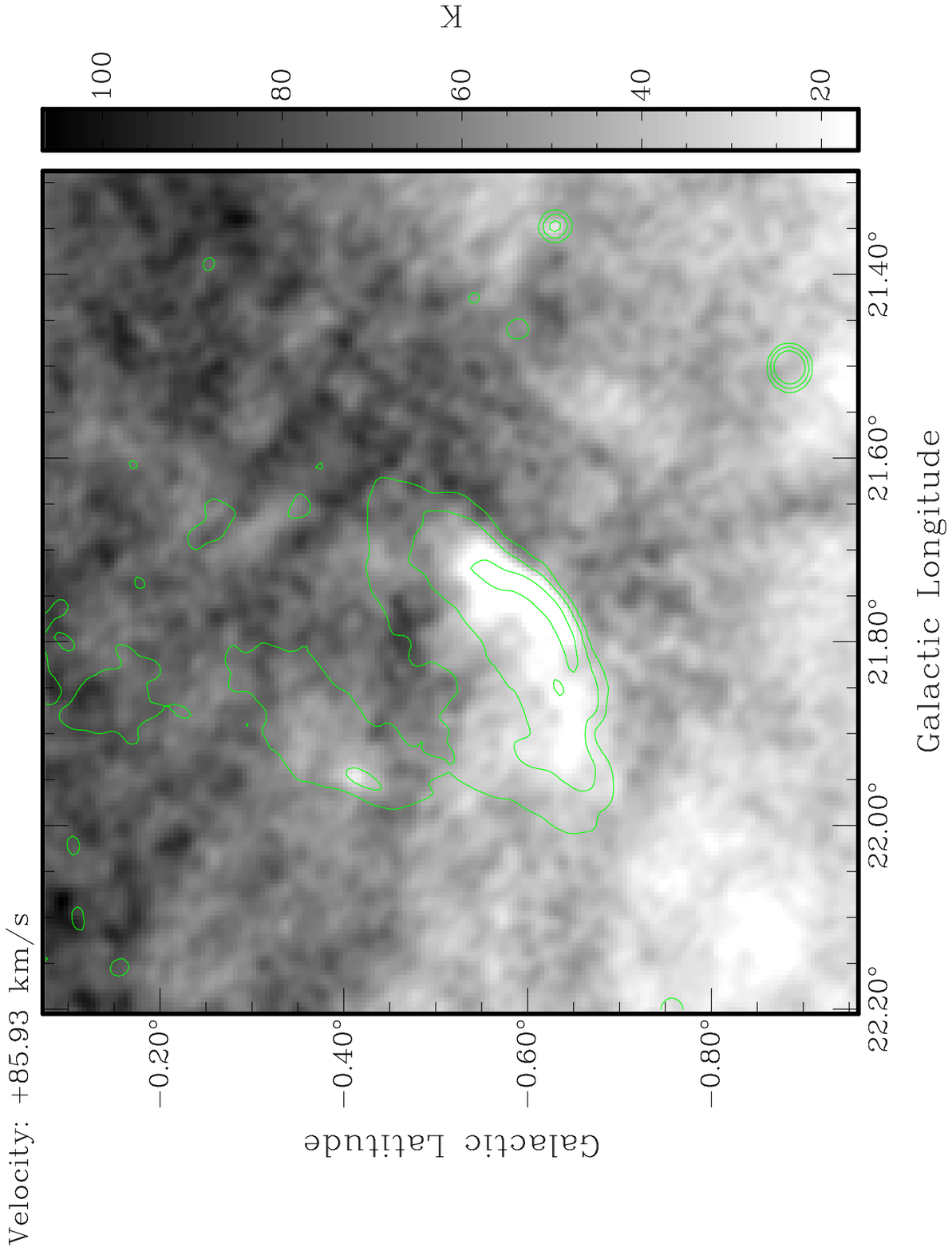}}
\put(-235, 210){\includegraphics{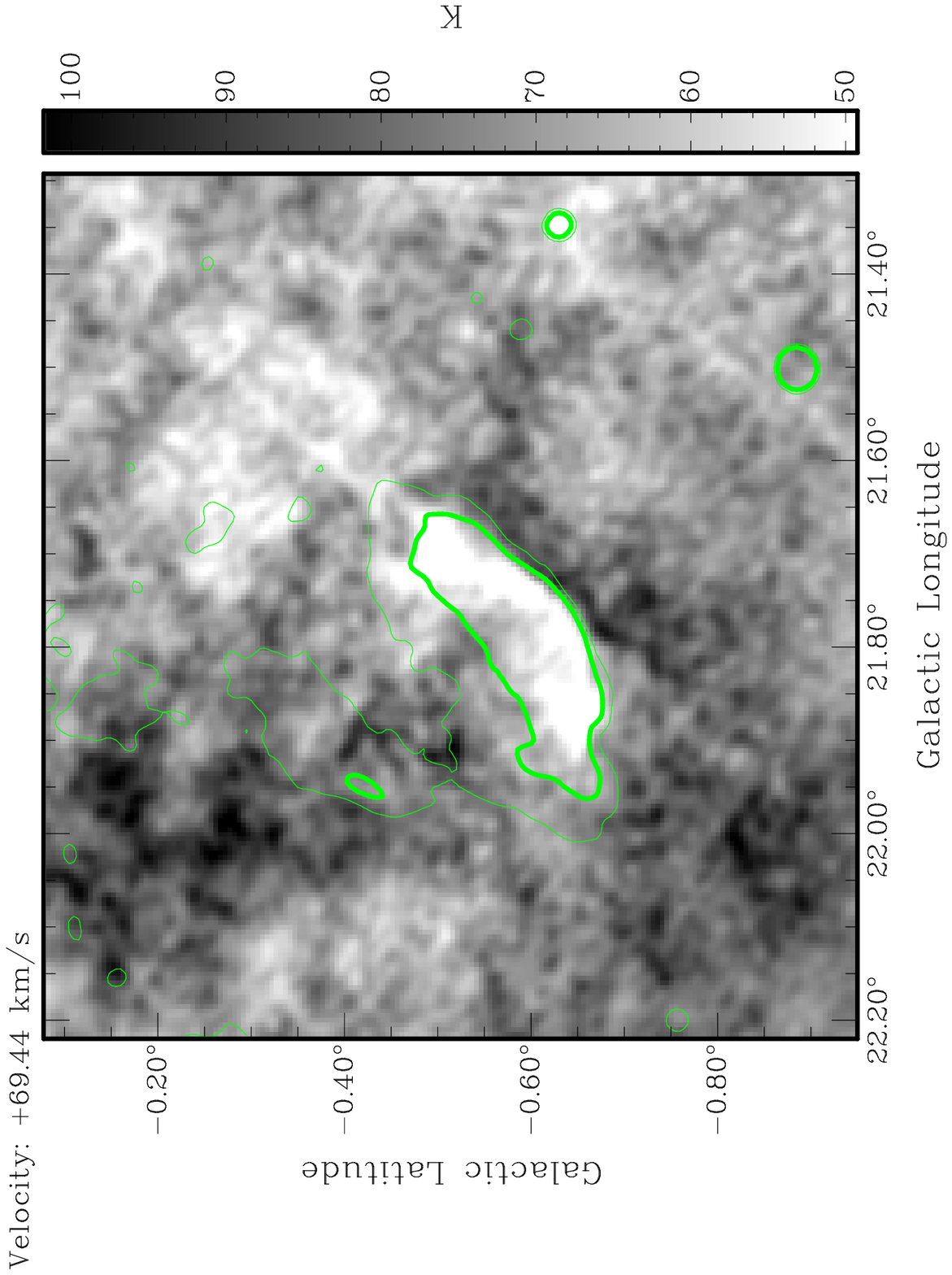}}
\put(80, -80){\includegraphics{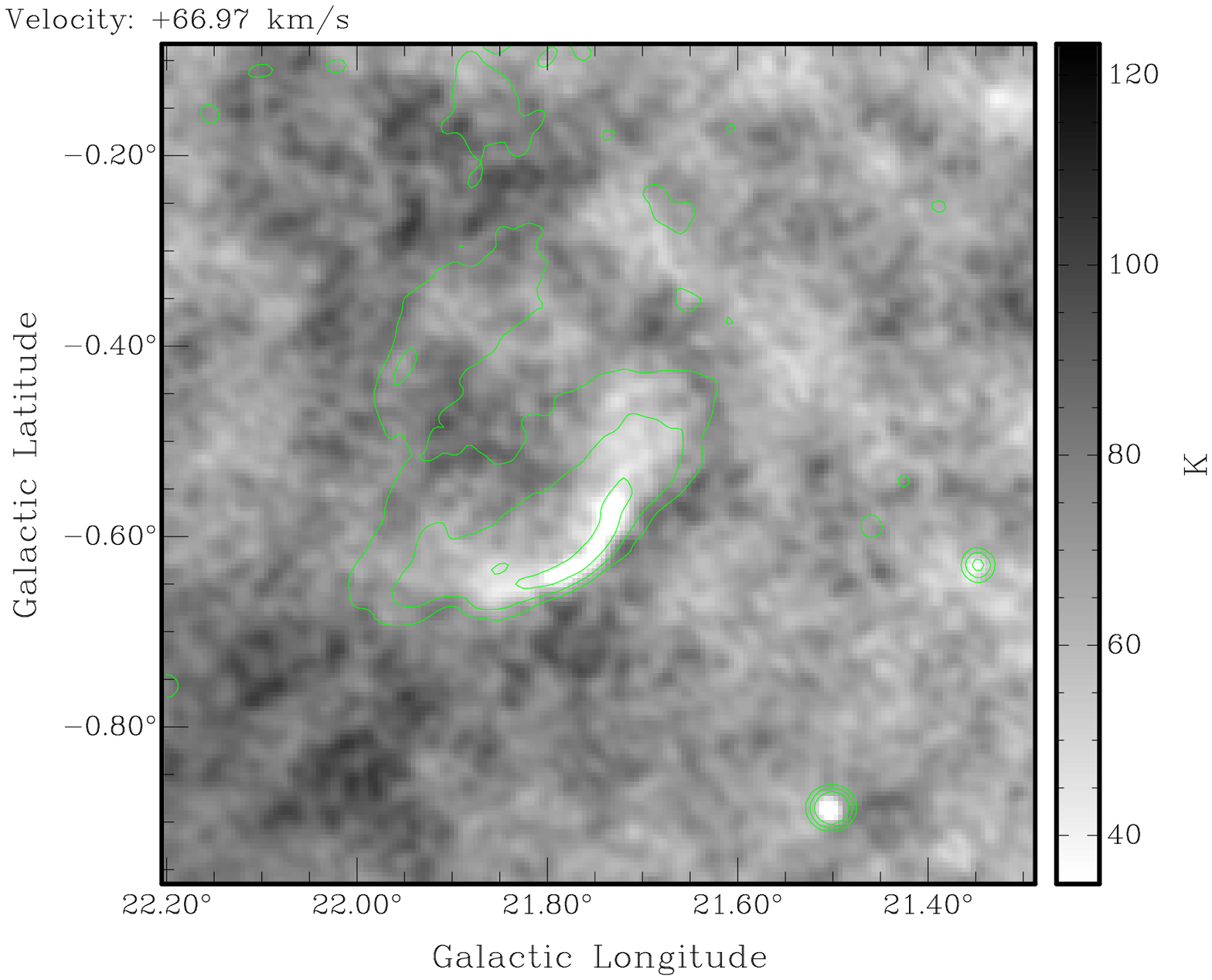}}

\end{picture}
\caption{First row: the 1420 MHz continuum image (left) and  a channel CO map with velocity of $\sim$ 86 km s$^{-1}$ (right), centered at [b=-0.5, l=21.75]; Kes 69 is the large extended object near center, G21.5-0.9 is inside box 3, PMN J1832-1035 is inside box 4. Second and third row: HI maps at velocities of $\sim$ 112 km s$^{-1}$ (close to the tangent point velocity in the direction towards Kes 69, middle left), 86 km s$^{-1}$ (middle right), 69 km s$^{-1}$ (lower left) and 67 km s$^{-1}$ (lower right), respectively. The first panel has superimposed contours (green: 25, 43, 75, 118, 145K) of the 1420 MHz continuum emission to show the SNRs, other panels have superimposed contours (green: 25, 50, 500 K).  In the first panel, the plus sign shows the site of the 1720 MHz OH maser at $\sim$ 70 km s$^{-1}$, the four boxes mark  regions where  HI and CO spectra are extracted.}
\end{figure*}

\begin{figure*}
\vspace{150mm}
\begin{picture}(120,120)
\put(-180,420){\includegraphics{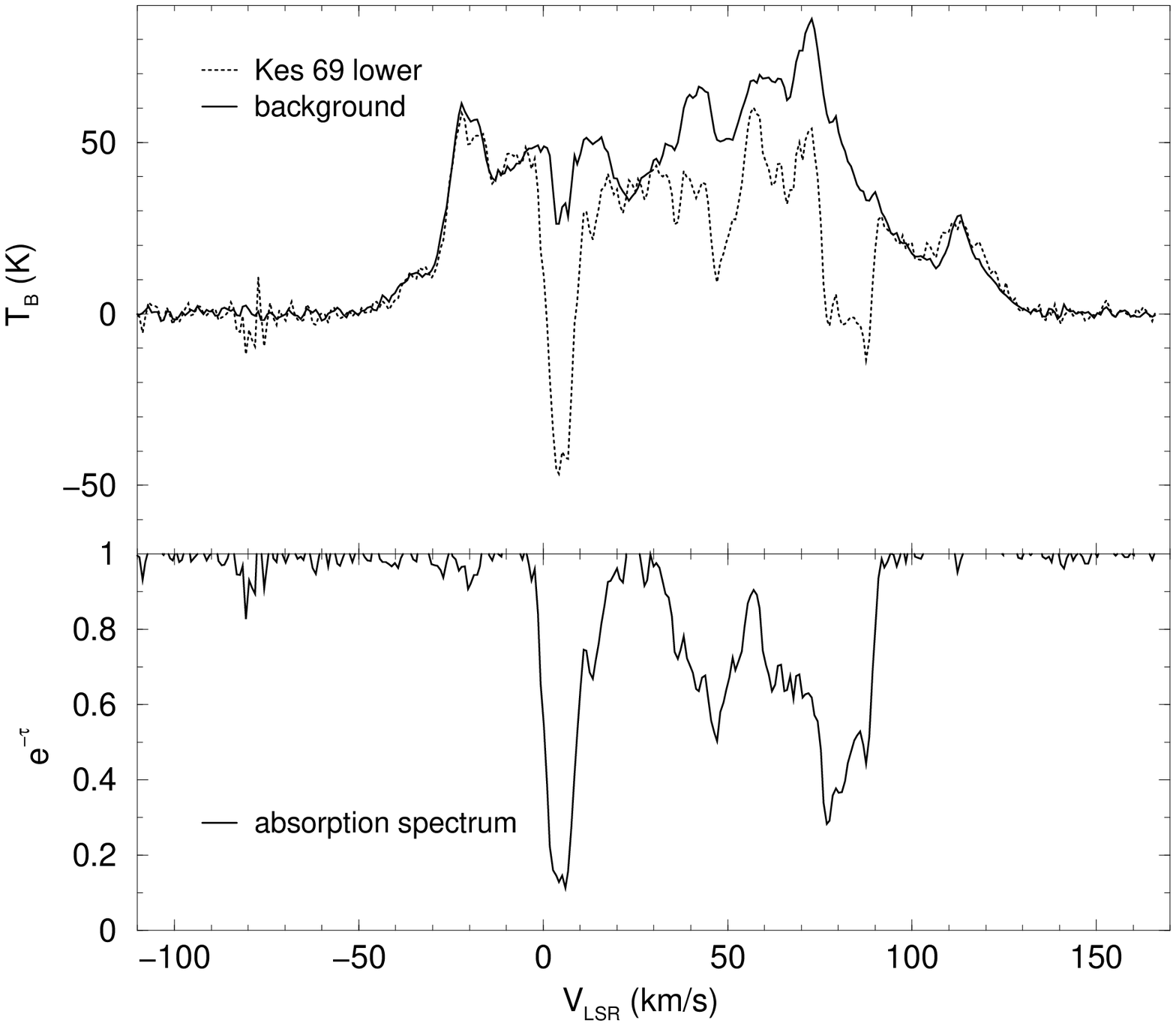}}
\put(60,560){\includegraphics{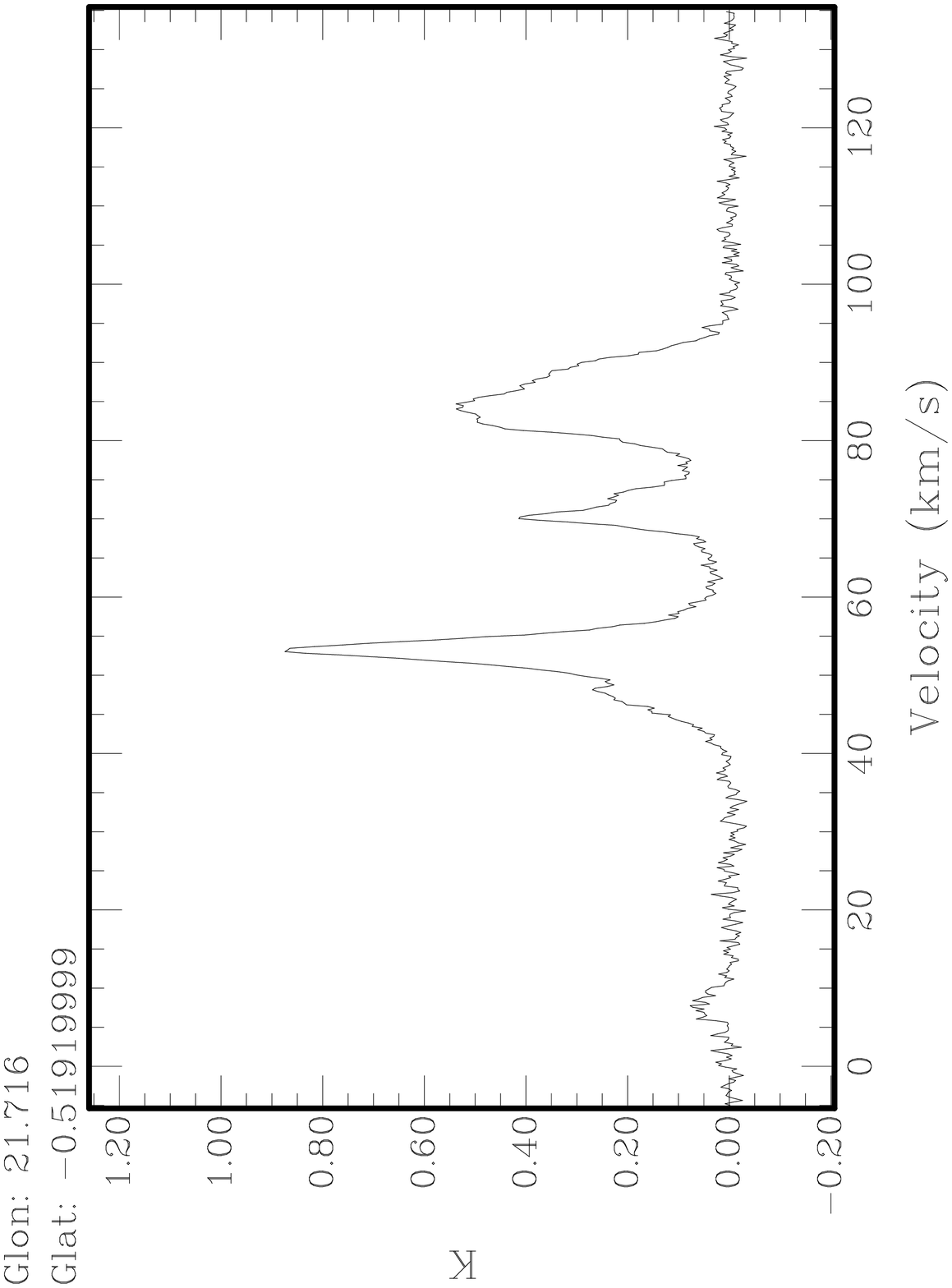}}
\put(-180,275){\includegraphics{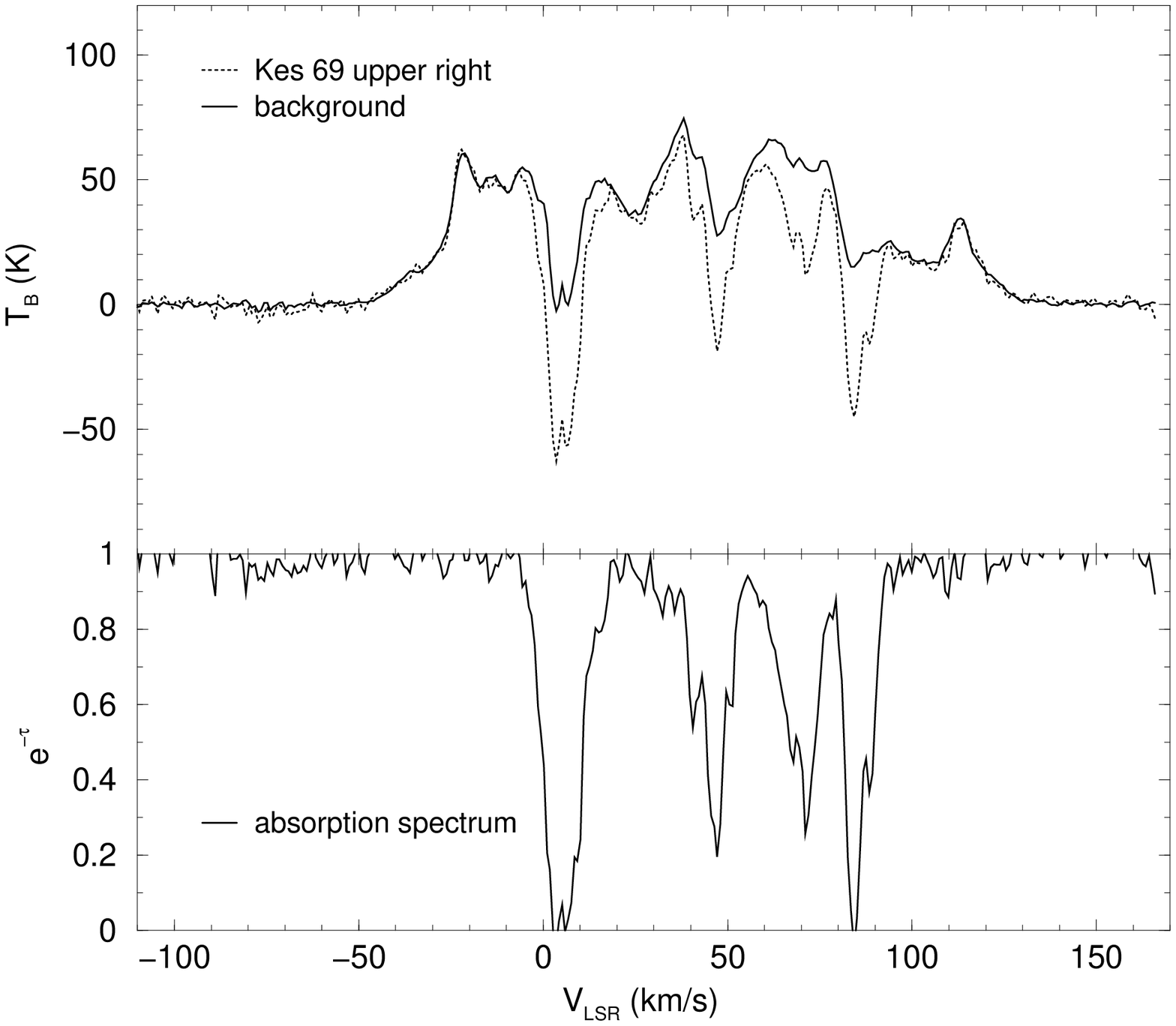}}
\put(60,415){\includegraphics{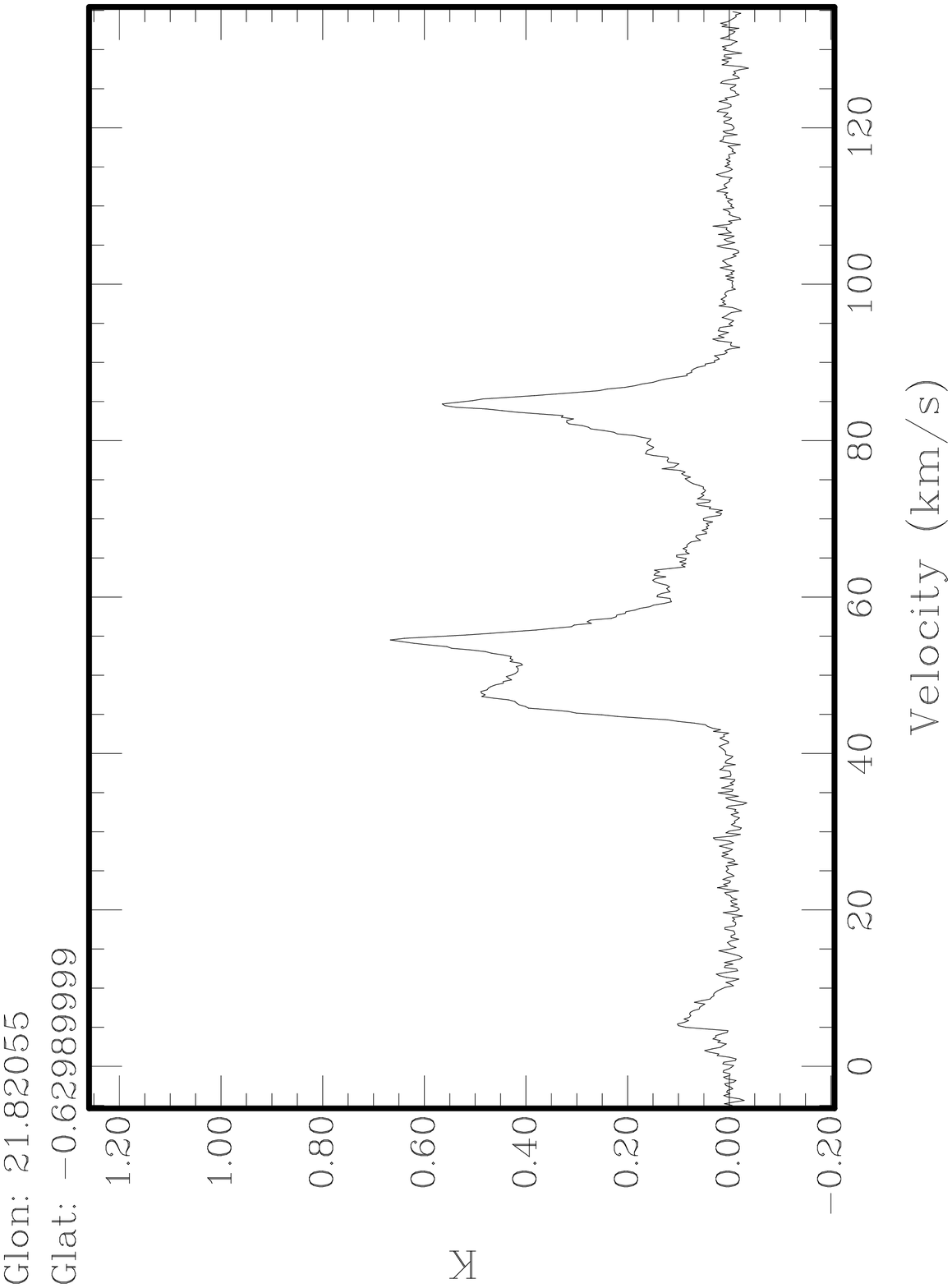}}
\put(-180,140){\includegraphics{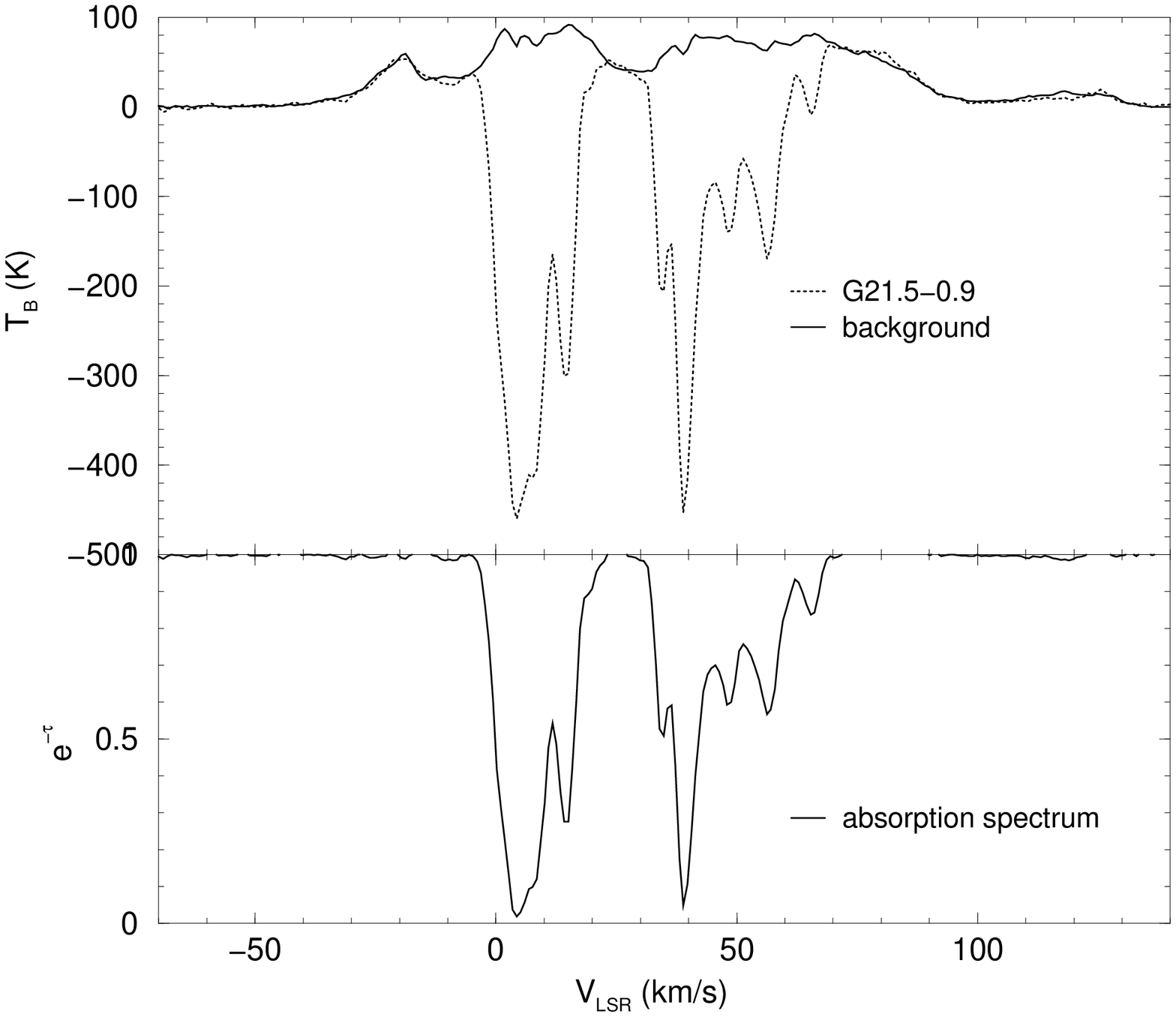}}
\put(60,280){\includegraphics{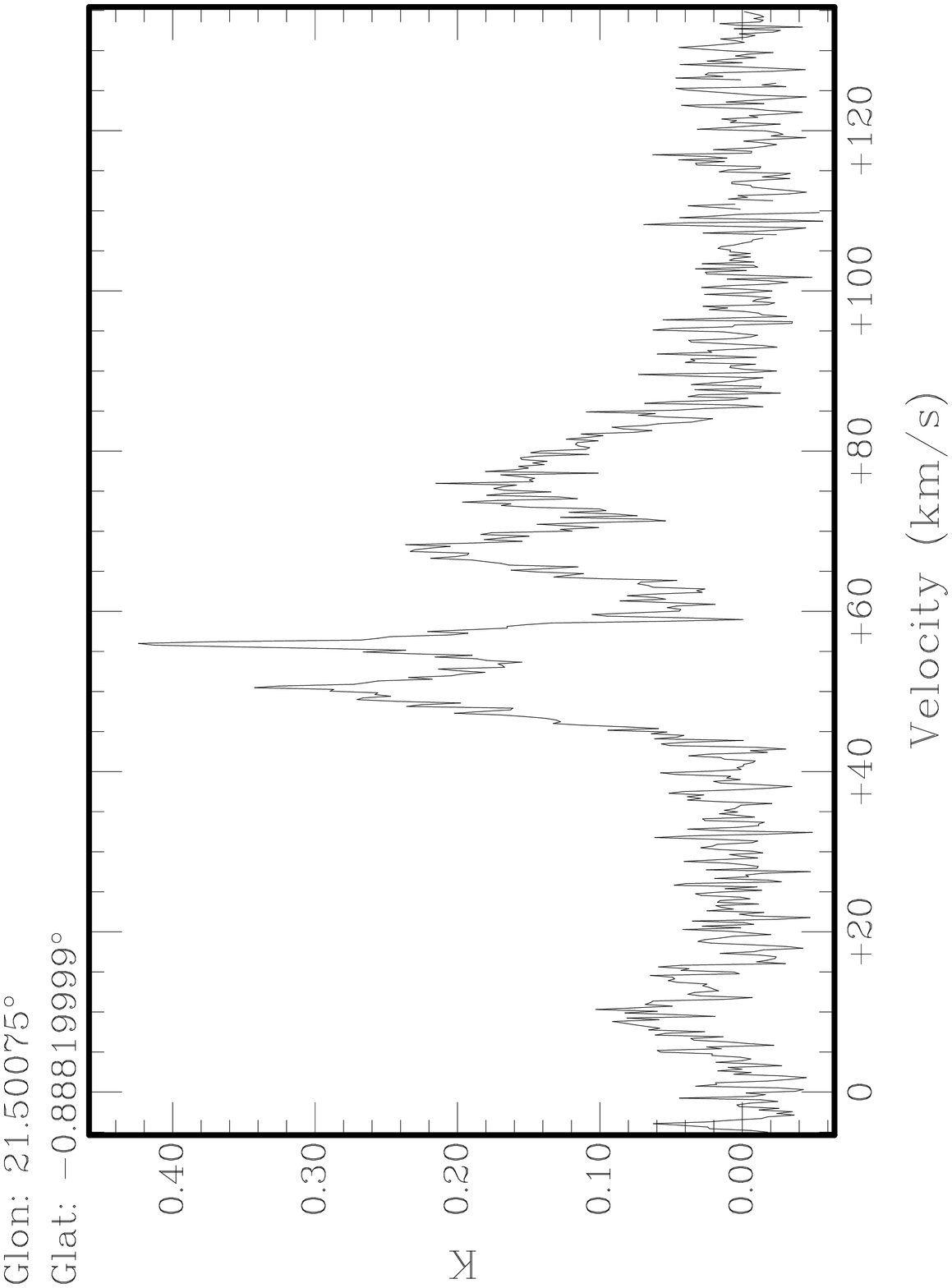}}
\put(-180,0){\includegraphics{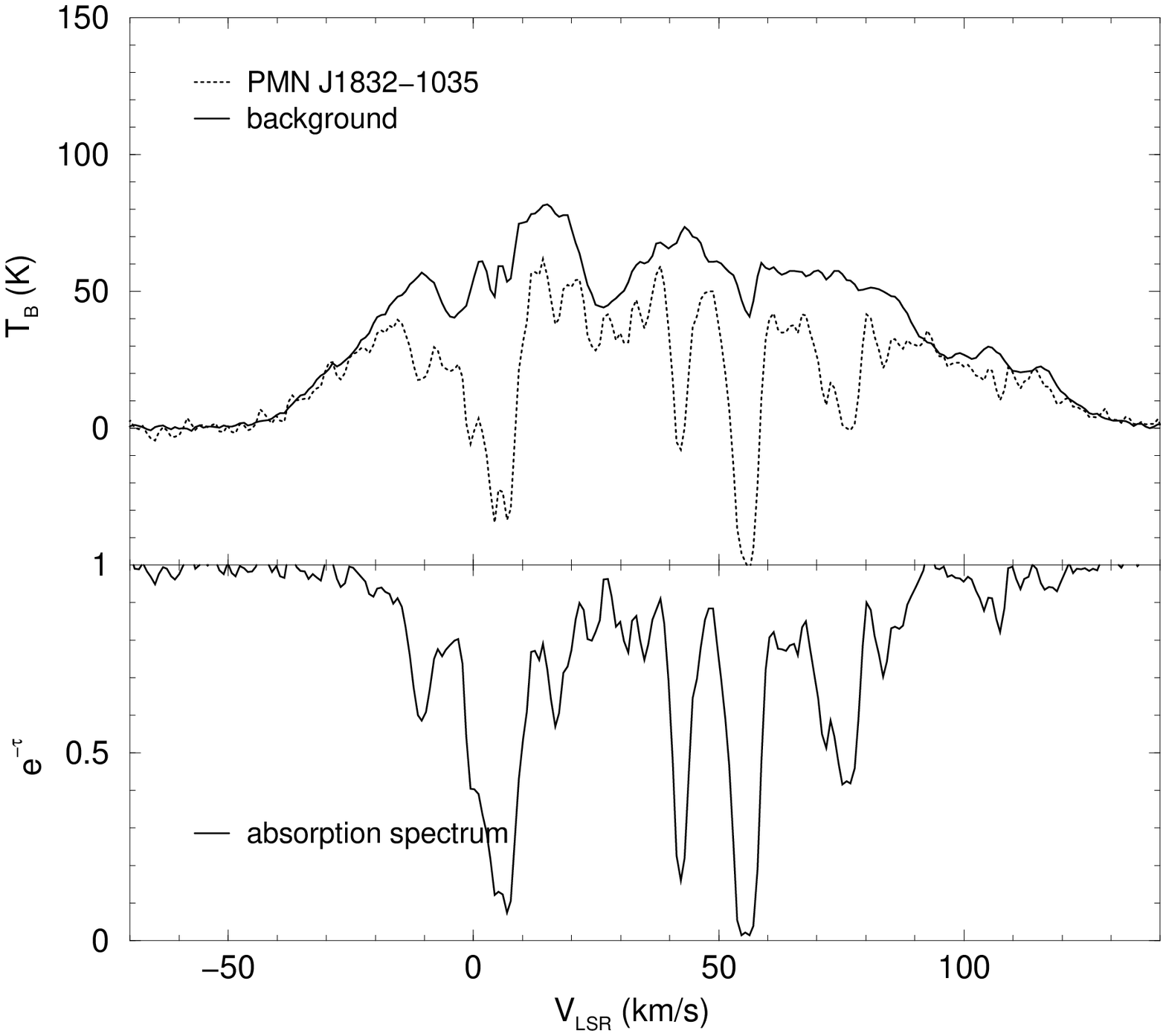}}
\put(60,140){\includegraphics{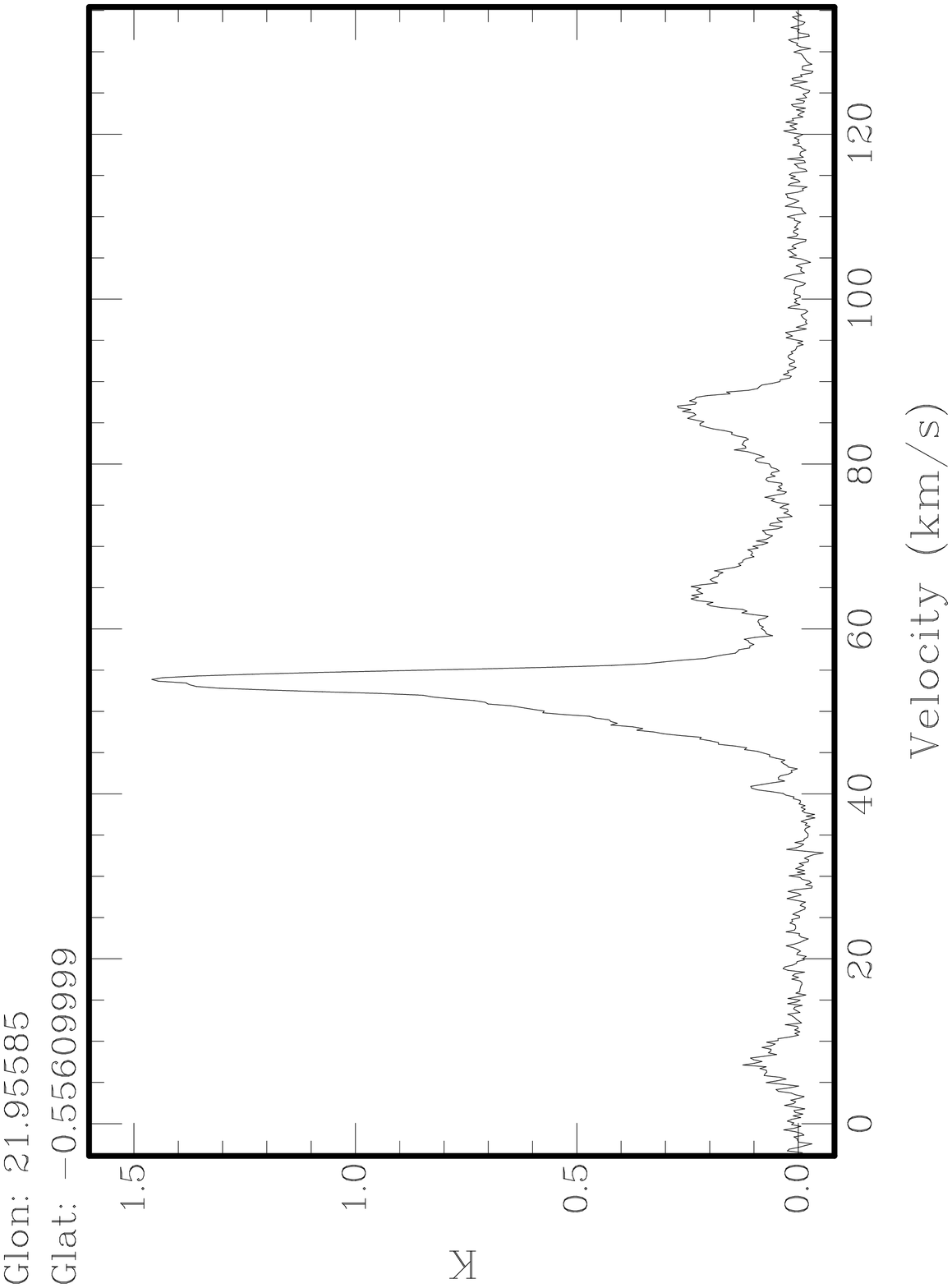}}
\end{picture}
\caption{Left: four HI emission and absorption spectra (from top to bottom), extracted from boxes 1, 2, 3 and 4 shown in Fig. 1a. Right: four CO emission spectra, extracted from boxes 1, 2, 3 and from OH maser site shown in Fig. 1a.}
\end{figure*}

\section*{Acknowledgements}
WWT and DAL acknowledge support from the Natural Sciences and Engineering Research Council of Canada. WWT appreciates support from the Natural Science Foundation of China.  
This publication makes use of molecular line data from the Boston University-FCRAO Galactic Ring Survey (GRS). The GRS is a joint project of Boston University and Five College Radio Astronomy Observatory, funded by the National Science Foundation.  
The NRAO is a facility of the National Science Foundation operated under cooperative agreement by Associated Universities, Inc.

\end{document}